\pdfoutput=1

\documentclass[aps,preprintnumbers,amsmath,amssymb,twocolumn, tightenlines,superscriptaddress,nofootinbib]{revtex4}
\usepackage{graphicx}
\usepackage{slashed}
\usepackage{lipsum}
\usepackage{tikz,mathpazo}
\usetikzlibrary{shapes.geometric, arrows}
\usepackage[none]{hyphenat}
\usepackage[english]{babel}

\begin{document}


\title{Rotating gluon system and confinement}

\author {Yin Jiang} 
\address{Physics Department, Beihang University, 37 Xueyuan Rd, Beijing 100191, China}
\address{Beihang Hangzhou Innovation Institute, Yuhang, Hangzhou, 310023, Chin}
\date{\today}

\begin{abstract}
In this work the non-abelian gauge theory is reformulated in a local inertial frame with the presence of a background rotation. With this new formalism the influence of the background rotation on the color deconfinement transition for a SU(2) pure gluon system. The KvBLL caloron, which is a color neutral and asymptotically nontrivial solution of Yang-Mills equation at finite temperature, is adopted to confine the color charges. With new solutions of the caloron's constituent particles, i.e. dyons, the semi-classical potential, which confines color charges, and the perturbative potential, induced by the Gaussian fluctuation, have been obtained for this system under rotation. By solving the critical temperature of confinement-deconfinement phase transition in different computation schemes, it is found that neither the rotational semi-classical potential nor Gaussian fluctuations can confine color charges more tightly when the rotation becomes faster. While only a stronger coupling constant is able to make the critical temperature increasing with angular velocity, as that indicated in lattice simulations. And it is also found with some particular sets of parameters, a non-monotonic dependence of the critical temperature will be obtained in the most physically realistic case, in which all the three factors are taken into account.
\end{abstract}
\maketitle

\section{Introduction}

Quark-Gluon-Plasma, which is generated in relativistic heavy-ion collisions, is believed to have the highest vorticity in the world. The globally averaged rotation speed and total angular momentum of the QGP can be detected and then estimated through polarization measurements of vector hadrons\cite{STAR:2017ckg, STAR:2018gyt, Xia:2018tes, Liang:2019pst, Ivanov:2019ern, Guo:2019joy, STAR:2020xbm, STAR:2021beb,  Ryu:2021lnx,  Tuchin:2021lxl, Deng:2021miw, Fu:2020oxj, Lei:2021mvp, Li:2021zwq}. Such measurements show encouraging qualitative agreement with the corresponding phenomenological simulations, especially the decreasing trend of the averaged vorticity and angular momentum's dependence on the collision energy \cite{Wei:2018zfb, Deng:2020ygd, Yi:2021unq,  Ivanov:2020wak, Rybalka:2018uzh, Wu:2019eyi, Yi:2021ryh}. However, a closer examination of the local vorticity reveals that the rotation-relevant physics may be far from clear\cite{Ayala:2019iin, Wang:2018sur, Wei:2021dib, Yang:2017sdk, Singh:2018bih, Li:2022pyw}. Although the quantum transport theory  has already characterized the dominant quantum effect of light quarks, which are approximate chiral fermions in QGP, there are still significant differences between simulations and experiments, such as the local vorticity distribution and global polarization of different hadrons\cite{Huang:2017pqe, Sheng:2021kfc, Sun:2016mvh, Chen:2020xsr, Abramchuk:2018jhd, Fukushima:2020ucl, Buzzegoli:2022dhw}. Hence it is indicated that there may be more non-trivial properties of the background QGP, which provides the vorticity of the system. The soft gluons, the primary constituents of QGP, are supposed to suffer from much more rotational effects than quarks due to their greater spin number. Therefore phase structure, particularly the confinement/deconfinement, of the background rotating gluon matter appears to be crucial for comprehending the transport mechanism within QGP by influencing the hydrodynamic transport properties. 


For a static pure gluon system, a potential scenario of the color confinement is to consider a topologically non-trivial background gluon configurations to confine charges. The KvBLL caloron, which has non-trivial expectation value of the Polyakov loop at spatial infinity (holonomy), is a good choice to reproduce the confinement/deconfinement transition at finite temperature\cite{Kraan:1998sn, Kraan:1998pm, Lee:1998bb, Diakonov:2005qa, Lawrence:2023woz, Escobar-Ruiz:2017uhx,   Diakonov:2002qw, Diakonov:2007nv, Schafer:1996wv, Kharzeev:2007jp}. In the framework of thermal field theory, the conventional approach is to compute the thermodynamic potential or free energy of the system. To the second order of quantum fluctuations, there are two components in the free energy. The first component originates from the background field and depends on the action of the selected background gluon configuration. This component, which is minimized at the non-trivial holonomy, serves as the confined part. The second component arises from Gaussian integration of quantum fluctuations \cite{Gross:1980br, Weiss:1980rj} which is formally like the free energy of free gluon gas. This always leads to trivial holonomy and thus deconfinement. As temperature increases, the confined part becomes weaker while deconfined part stronger. Consequently, deconfinement occurs at the critical temperature. This entire scenario can also be translated into the rotational case. In principle, three factors will modified the corresponding free energy. The first one pertains to the rotational impacts on the background gauge field. This can be computed by explicitly solving the Yang-Mills equation with rotation corrections. The second involves computing the Gaussian integration while considering finite-size and angular momentum polarization effects. Lastly, the correlation between the coupling constant and rotation speed may significantly influence the color deconfinement transition. In recent studies there are several works on deconfinement phase transition by considering some of these three factors\cite{Fujimoto:2021xix, Chen:2020ath, Braga:2022yfe, Chen:2022smf} to explain the unexpected critical temperature behavior suggested by lattice QCD \cite{Braguta:2021ucr, Braguta:2021jgn, Yang:2023vsw}. In this work we opt for the KvBLL caloron as the confined vacuum and take into account all three factors. 
Our results indicate that only a stronger coupling constant can confine color charges. While both the QCD vacuum and quantum fluctuations contribute to an increased likelihood of deconfinement as rotation speed rises. The interplay between these two opposing trends ultimately may result in a non-monotonic relationship between the critical temperature and rotation speed.


In a rotating case the real velocity introduces new imaginary terms to the gluon's Lagrangian density at finite temperature. This leads to a novel kind of sign problem in the lattice QCD simulation. According to studies on such complex Lagrangian density problems\cite{Witten:2010zr, Aarts:2014nxa, Alexanian:2008kd, Cherman:2014ofa, Dunne:2015eaa, Nishimura:2014rxa}, it does not pose a problem for our study since only a semi-classical background and Gaussian fluctuations are considered within our framework. Consequently, we can safely retain the imaginary component as is or initiate with an imaginary velocity and perform analytical continuation by simply replacing it with a real one eventually.

\section{Gauge field in local inertial frame}
Consider a globally rotating system of pure gluons with an constant angular velocity $\vec{\omega}$, in the local rest frame the rotation is equivalent to a curved metrics of the space-time as follows
\begin{eqnarray}
g_{00}=1-\vec{v}^2,\ 
g_{m n}=-\delta_{m n},\ 
g_{0 m}=g_{m 0}=-v_m.
\end{eqnarray}
where $v_m=(\vec{\omega}\times\vec{x})_m=\epsilon_{m n l}\omega_n x^l$ and $m, n, l=1, 2, 3$. And $v_m$ are different components of the physical linear velocity vector, i.e. we do not distinguish between $v_m$ and $v^m$ and only use $v_m$ in the following. In order to work in the local {\it inertial} frame, where more easily Dirac spinors introduced in future, all the general vectors, indexed by $\mu, \nu, \lambda, ...=0, 1, 2, 3$, $T^\mu$ and tensors in such a curved space-time should be projected to the local Minkovski basis with the help of vierbein fields $e_\mu^a$ in the way of $T^a=e_\mu^a T^\mu$, where the $a, b, c,...$ indexed $T^a$ is the corresponding component in the local inertial frame. The two vierbein fields satisfy their definitions 
\begin{eqnarray}
e_\mu^a e_\nu^b \eta_{a b}=g_{\mu\nu},\\
\xi^\mu_a \xi^\nu_b g_{\mu\nu}=\eta_{a b}.
\end{eqnarray}
where $\eta_{ab}=diag(1, -1, -1, -1)$ is the metrics of Minkowski space-time. Obviously they are used to project vectors in curved/flat to flat/curved space-time respectively. The choice of vierbein fields is not unique. And in this work we choose a simple form as
\begin{eqnarray}
e_0^0=1,
e_0^m=v_m;\\
\xi_\mu^\mu=1,
\xi_0^\mu=-v_m.
\end{eqnarray}
With the result of the covariant derivative
\begin{eqnarray}
\nabla_b A_a=\xi^\mu_b D_\mu A_a=\xi^\mu_b (\partial_\mu A_a+\xi_a^\nu e^c_{\nu;\mu}A_c)
\end{eqnarray}
we obtain the field strength tensor in the local inertial frame as
\begin{eqnarray}
F^i_{ab}=\nabla_a A^i_b-\nabla_b A^i_a+g f^{i j k}A_a^j A_b^k.
\end{eqnarray}
In the following we will use $i, j, k=1, 2, 3$ as color indices and $a, b, c=0, 1, 2, 3$ the space-time indices and $m, n, l=1, 2, 3$ the corresponding spatial indices in the inertial frame. The gauge symmetry is preserved by noticing its relation with the field strength tensor in the curved space-time as
\begin{eqnarray}
F^i_{ab}=\xi_a^\mu\xi_b^\nu F_{\mu\nu}^i.
\end{eqnarray}
And the gauge transformation is also modified by considering the projection as 
\begin{eqnarray}
\label{gauge}
A_a \rightarrow U A_a U^\dagger+ig^{-1}U\nabla_a U^\dagger.
\end{eqnarray}
Because the gauge group element $U$ is only a space-time scalar, the covariant derivative of it is reduced as $\nabla_a U^\dagger=\xi_a^\mu \partial_\mu U^\dagger$. 
And the explicit form of the field strength tensor in local inertial frame is obtained as
\begin{eqnarray}
\label{fuv}
&&F_{0m}^i=\partial_0 A_m^i-\partial_m A_0^i+A_n^i \partial_n v_m -v_n\partial_n A_m^i +g f^{i j k}A_0^j A_m^k;\nonumber\\
&&F_{m n}^i=\partial_m A_n^i-\partial_n A_m^i+g f^{i j k}A_m^j A_n^k.
\end{eqnarray}
In order to study the critical temperature of deconfinement transition at finite angular velocity we perform the usual Wick rotation on the action into the Euclindean space-time by replacing $x_0$ with $-i x_4$ and $A_0$ with $i A_4$. It is clear that all the temporal terms become imaginary while the others stay real in $F_{0 m}$. This make all the velocity relevant terms are imaginary in the gauge field Lagrangian density which causes a new kind of sign problem in the corresponding lattice simulation. Nevertheless, it does no harm to semi-classical solutions except making the gauge field complex. We will use imaginary angular velocity to have a real formalism in the following computation. Eventually we replace the imaginary velocity with a real one and obtain all the numerical results. For the Gaussian fluctuation part we always use a real angular velocity to compute the perturbative potential.

Because the spatial parts of $F_{ab}$ in Eq.\ref{fuv} are not modified by rotation, we attempt to find a semi-classical solution only whose temporal component $A_4$ has acquired an extra term $\delta A_4$ to preserve the self-dual and anti-self-dual relation with presence of rotation
\begin{eqnarray}
\label{dual}
F_{4m}=\pm\epsilon_{mnk}F_{nk}
\end{eqnarray}
where $m, n, k=1, 2, 3$. And it is known that such relations are equivalent to the standard Yang-Mills equation because of the Bogomol’nyi inequality in static case. And it can also be  explicitly checked that all the configurations in next section are solutions to the rotational Yang-Mills equation.  
Besides the gauge transformation and semi-classical solution, Polyakov loop's definition is supposed to be modified by the rotation as well. The traditional Polyakov loop is defined as
\begin{eqnarray}
L=\mathcal{P} e^{i \int_0^\beta dx_4 A_4}    
\end{eqnarray}
because the following property is satisfied under a gauge transformation U
\begin{eqnarray}
\label{link}
L^U=U(\beta, \vec{x})LU^\dagger(0, \vec{x}).
\end{eqnarray}
In the imaginary-time thermal field theory, the $x_4$-dependent gauge transformation should satisfy the periodic condition along the imaginary time axis up to an element $z$ of the center of the gauge group
\begin{eqnarray}
U(\beta, \vec{x})=zU(0, \vec{x}).
\end{eqnarray}
Once the $Tr\langle L\rangle$ is nonzero, the center symmetry is broken, which means the occurrence of the deconfinment transition. However the property in Eq.\ref{link} is preserved basing on the explicit form of the gauge transformation. Once the transformation is altered as Eq.\ref{gauge} the property in Eq.\ref{link} will no longer be valid. This means the definition of Polyakov loop should be modified in the way of constructing a new combination of guage fields to make the property in Eq. \ref{link} and the criteria of center symmetry established again. A potential construction could be 
\begin{eqnarray}
\label{pl}
L=\mathcal{P} e^{i \int_0^\beta dx_4 (A_4+v_m A_m)},
\end{eqnarray}
because the combination $\Phi=A_4+v_m A_m$ transforms as the $A_4$ in the static system, that is 
\begin{eqnarray}
A^U_4+v_m A^U_m=U(A_4+v_m A_m)U\dagger +i U\partial_4 U^\dagger
\end{eqnarray}
Although the spatial component $A_m$ usually approaches to zero at spatial infinity, it is also possible to provide a nonzero correction to the Polyakov loop for a proper field configuration. Because the linear velocity $v_m$ is linear with spatial distance.

\section{Semi-classical solutions}
Now we focus on the case of SU(2) gauge group. According to Eq.\ref{fuv} we rewrite the color electric field as
\begin{eqnarray}
F_{4m}=\pm\epsilon_{mnk}F_{nk}+\delta F_{4m}.
\end{eqnarray}
Here $F_{ab}$ represents the field strength tensor of a certain solution in the static case. Hence the extra terms should satisfy
\begin{eqnarray}
\label{a4}
\delta F_{4m}^i&&=-\partial_m \delta A_4^i+A_n^i \partial_n v_m -v_n\partial_n A_m^i +g \epsilon^{i j k}\delta A_4^j A_m^k\nonumber\\
&&=0
\end{eqnarray}
for a self or anti-self dual solution. And the two $A^i$ terms can be reduced in the cylindrical coordinate $(\varrho, \phi, z)$ as follows 
\begin{eqnarray}
\label{a4p}
&&A_n^i \partial_n v_m -v_n\partial_n A_m^i
=(\vec{A^i}\cdot\nabla)\vec{v}-(\vec{v}\cdot\nabla)\vec{A^i}\nonumber\\
&&=-\omega(\partial_\phi A_\varrho\hat{\varrho}+\partial_\phi A_\phi\hat{\phi}+\partial_\phi A_z\hat{z})
\end{eqnarray}
where $\vec{v}=\omega \varrho \hat{\phi}$ has been used.
With a given solution $A_a$ in static system once a $\delta A_4$ can be found to satisfy Eq.\ref{a4}, a corresponding solution in rotating system is obtained as $A^{(\omega)}=(A_4+\delta A_4, \vec{A})$. It is believed that the KvBLL caloron is a good candidate for the confined QCD vacuum. Because this color neutral solution satisfies the periodic condition along the $x_4$ direction and has non-trivial asymptotic behavior with the help of construction by superposing of two dyons(named M and L). In this work we consider the dilute caloron and dyon well-separated limit. 
In such limit the system can be treated as a many-body system of the same amount of M and L dyons with no interactions. 
Hence we set the background gluon field as dyon and solve Eq.\ref{a4}. 

There are two types of dyons(self-dual) in the SU(2) case, i.e. M and L dyon. Each type has its anti particle(anti-self-dual), named $\bar{\text M}$ and $\bar{\text L}$ respectively. Each dyon carries net color electric and magnetic charges which shown in Table \ref{dnlist}. In hedge-hog gauge the BPS M(+) and $\bar{\text M}$(-) dyons are
\begin{eqnarray}
\label{mdyon}
A^i_4=\pm n_i (\frac{1}{r}-\rho\ \coth(\rho r))\\
A^i_m=\epsilon_{i m k}n_k(\frac{1}{r}-\rho\ \text{csch}(\rho r)).
\end{eqnarray}
where $n_i=x_i/r$ and $r=|\vec{x}|$. And the $\rho$ is a free parameter of the solution and its inverse $\rho^{-1}$ characterizes the size of a dyon. One may notice that the color index $i$ binding with the spatial direction $n_i$ and $\omega_i$ here and in following. It is the very particular property of such topologically non-trivial solutions which mapping $S^1\otimes R^3$ space-time to the $su(2)$ color space. And $L$ and $\bar{L}$ dyons are obtained by replacing $\rho$ with $2\pi T-\rho$. The asymptotic behavior of temporal component of dyons read
\begin{eqnarray}
&&A^{{\text M}, \bar{\text M}}_4\rightarrow \mp \rho n_m \frac{\tau_m}{2} \\
&&A^{{\text L}, \bar{\text L}}_4\rightarrow \mp (2\pi T-\rho) n_m \frac{\tau_m}{2}.
\end{eqnarray}
It is determined by minimizing the free energy of the system. Obviously it is equivalent to the order parameter, i.e. the Polyakov loop, of the deconfinement transition.

In the rotational case the additional term $\delta A^i_4$ of dyons is obtained by solving Eq.\ref{a4} 
\begin{eqnarray}
\delta A_4^i=\omega_i. 
\end{eqnarray}
Here the gauge field $A_m^i$ in Eq.\ref{a4} is the solution of dyons in static case in Eq.\ref{mdyon}.
This means in a rotating system the asymptotic behavior of BPS dyons become
\begin{eqnarray}
&&A^{M, \bar{M}}_4\rightarrow (\omega_m \mp \rho n_m) \frac{\tau_m}{2} \\
&&A^{L, \bar{L}}_4\rightarrow (\omega_m \mp (2\pi T-\rho) n_m) \frac{\tau_m}{2}.
\end{eqnarray}
The corresponding eigenvalues are $\pm \sqrt{\omega^2+\rho^2\mp \vec{n}\cdot\vec{\omega}}$ for M, $\bar{\text M}$ dyons and  $\pm \sqrt{\omega^2+(2\pi T-\rho)^2\mp \vec{n}\cdot\vec{\omega}}$ for L, $\bar{\text L}$ dyons. This result seems to indicate the Polyakov loop should be space-directional dependent. However it is not true because a single dyon can not serve as a color-neutral background field.
Dyons are usually {\it combed}(transformed to string gauge) to a chosen direction, such as the direction of angular velocity $\vec{\omega}$ which means we choose $\omega_i\tau_i=\omega \tau_3$. The corresponding gauge transformation is 
\begin{eqnarray}
&&S_+=e^{-i \frac{\phi}{2}\tau_3}e^{i \frac{\theta}{2}\tau_2}e^{i \frac{\phi}{2}\tau_3}\\
&&S_-=e^{i \frac{\phi}{2}\tau_3}e^{i \frac{\pi-\theta}{2}\tau_2}e^{i \frac{\phi}{2}\tau_3}
\end{eqnarray}
for M and $\bar{\text M}$ dyons respectively.
And L and $\bar{\text L}$ are transformed by an additional transformation
\begin{eqnarray}
U=e^{i \pi \tau_2/2}e^{-i\pi T x_4 \tau_3}
\end{eqnarray}
In this way the L mand M dyons get the same asymptotic behavior. And this makes it is possible to construct the caloron solution with the standard Atiyah--Hitchin--Drinfeld--Manin (ADHM) formalism.
The properties of different dyons, including action, non-zero component($A_4^3$) at spatial infinty, exponential factor $\Phi$ in Polyakov loop and color electric($q_E$) and magnetic($q_M$) charges, are summarized in Table \ref{dnlist}.
In the table $\rho_{M, L}$ are the corresponding parameter $\rho$ in the solution of M and L dyon. 

\begin{table}[ht]
\label{clist}
\centering
\begin{tabular}{||c c c c c c||} 
 \hline
 Dyon & Action & $A_4^3(\infty)$ & $\Phi(\infty)$ & $q_E$ & $q_M$ \\ [0.5ex] 
 \hline\hline
 $M$ & $\rho_M$ & $\rho_M-\omega$ &
 $\rho_M-\omega n_3$ & + & + \\ 
 \hline
 $\bar{M}$ & $\rho_M$ & $\rho_{\bar M}+\omega$ & 
 $\rho_{\bar M}+\omega n_3$ & + & - \\
 \hline
 $L$ & $2\pi T-\rho_L$ & $\rho_L+\omega$ &  
 $\rho_L+\omega n_3$ &- & - \\
 \hline
 $\bar{L}$ & $2\pi T-\rho_{\bar L}$ & $\rho_{\bar L}-\omega$ &
 $\rho_{\bar L}-\omega n_3$ &- & + \\
 \hline
\end{tabular}
\caption{Properties of SU(2) dyons in rotating system. The actions are shown in unit of $8\pi^2/g^2(2\pi T)^{-1}$.}
\label{dnlist}
\end{table}
From the table one can find that in the stringy gauge the $\delta A_4$ are transformed to $\mp \omega \tau_3/2$ for M/L dyon and $\pm \omega$ for $\bar{\text M} / \bar{\text L}$ dyon. These correction terms are very like the rotational polarization energies $n\omega$ acquired by different angular momentum states. Hence we name the $\delta A_4$ as the polarization term and the other terms of $A_a$ as normal terms in this work. Clearly the table also shows although the background rotation has changed the gauge field profile of dyons, the corresponding actions are still the same as those in the static case, i.e. which depends on the parameter $\rho$. This is a very important property because the semi-classical free energy of a single dyon($\rho=\rho_D$) depends on its action, that is
\begin{eqnarray}
\label{lam}
ln Z_D&&=2V |\rho_D|^3 \exp(-\frac{8\pi^2}{g^2}\frac{\rho_D}{2\pi T})\nonumber\\
&&=2V\rho_D^3\left(\frac{\Lambda}{\pi T}\right)^{\frac{22\rho_D}{6\pi T}}
\end{eqnarray}
Here the second equation has defined the parameter $\Lambda$ as $(\frac{\Lambda}{\pi T})^{22/3}=e^{-8\pi^2/g^2}$. 

\begin{figure}[ht]
    \centering
    \includegraphics[width=8 cm]{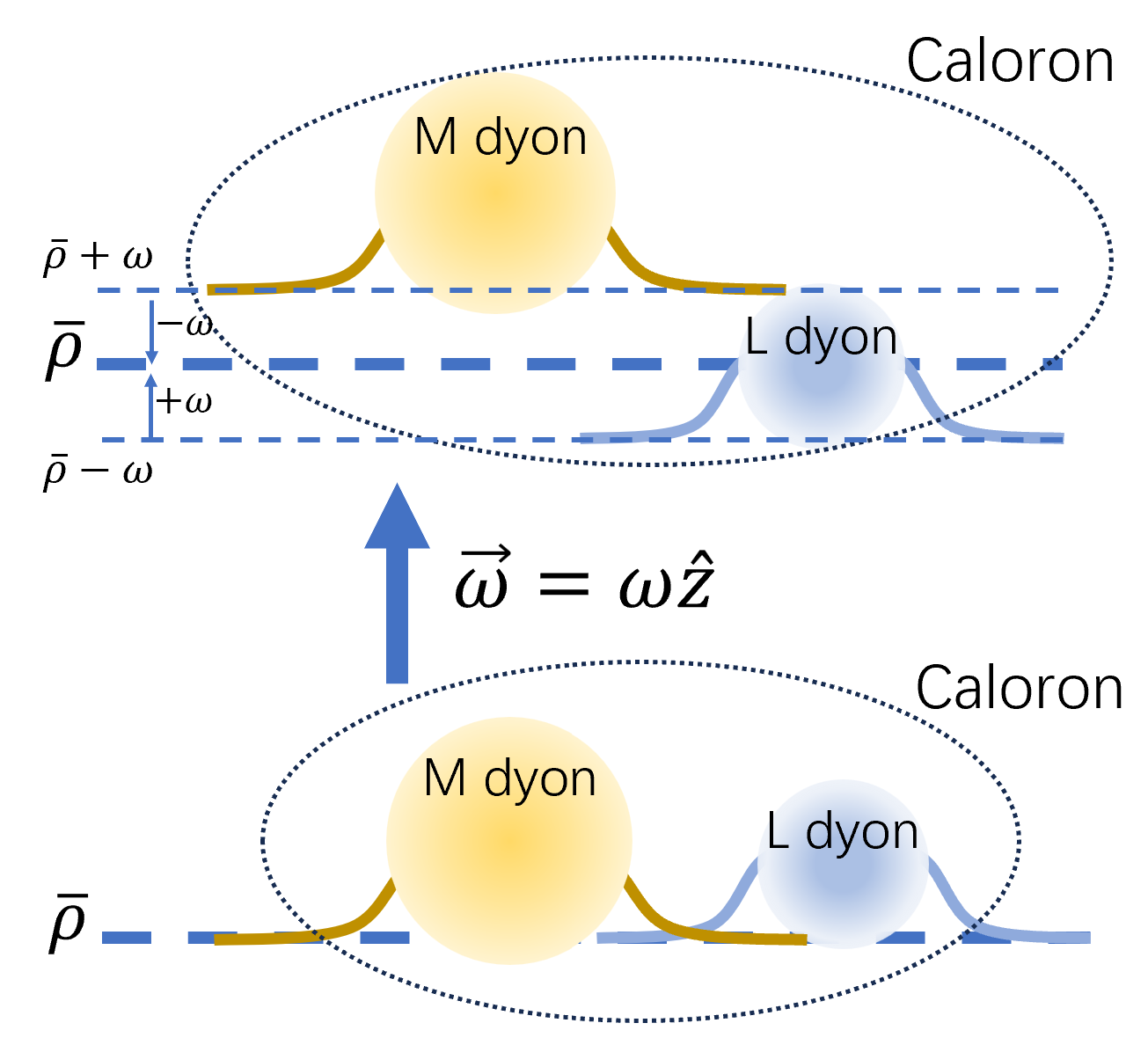}
    \caption{The critical deconfinement temperature as function of real(decrease) and imaginary(increase) angular velocity by considering the constant coupling and taking the rotational perturbative free energy.}
    \label{fig0}
\end{figure}

Although dyons have non-trivial asymptotic behavior at spatial infinity, one can not explain the confinement using a single dyon. Because the vacuum is supposed to be neutrally charged. At finite temperature a good choice is KvBLL caloron, which can be treated as superposition of a pair of M and L dyon with the {\it same} asymptotic behavior by the ADHM construction. 
And the semi-classical potential of caloron, without dyon-dyon interaction, is just the summation of actions of M and L dyon. In the region far from the core of M and L dyon the non-zero components of caloron is 
\begin{eqnarray}
&&A_4^{caloron}=\frac{\tau_3}{2}(\bar\rho+\frac{1}{r}+\frac{1}{s})\\
&&A_\phi^{caloron}=-\frac{\tau_3}{2}(\frac{1}{r}+\frac{1}{s})\sqrt{\frac{(r_{LM}-r+s)(r_{LM}+r-s)}{(r_{LM}+r+s)(r+s-r_{LM})}}\nonumber
\end{eqnarray}
where the $r$ and $s$ are distances from $\vec{x}$ to the center of M and L dyon and $\bar\rho$ the asymptotic value of caloron. And $r_{LM}$ is the distance between M and L dyon's center. In this work we assume the centers of M and L dyon are on the $x_3$ axis at $z_M$ and $z_L$ respectively for simplicity. Hence $r=\sqrt{x_1^2+x_2^2+(x_3-z_M)^2}$ and $s=\sqrt{x_1^2+x_2^2+(x_3-z_L)^2}$ and $r_{LM}=|z_L-z_M|$. Obviously neither the $A^{caloron}$ nor the $\vec{v}$ depends on azimuthal angle $\phi$. This means for caloron the rotation correction $\delta A_4=0$ according to Eq.\ref{a4p}. And at spatial infinity the $v_m A_m^{caloron}$ approaches zero which means the Polyakov loop is still $Tr(L)=2cos(\beta\bar\rho/2)$ as in the static case. Hence the composition picture of the caloron can be illustrated in the Fig. \ref{fig0}. That is, M and L dyon, whose normal part approaches $\bar\rho \pm \omega$ respectively, will be pinched together to $\bar\rho$ at infinite distance by their corresponding "polarization" terms $\mp \omega$ induced by the rotation. Thus the caloron can be constructed because of the same asymptotic behavior for both M and L dyons. More detailed construction and semi-classical potential computation are as follows. 

We consider a caloron whose asymptotic behavior is $\bar{\rho}$, that will not be influenced by the rotation and stay as 
\begin{eqnarray}
A_4^{caloron}(r\rightarrow +\infty)= {\bar \rho}\frac{\tau_3}{2}
\end{eqnarray}
In order to extract the semi-classical potential of the system, the two rotating constituent dyons' asymptotic behavior should be identified. Because the two dyons are supposed to have the same asymptotic behavior, according to Table \ref{dnlist}, the constraint means
\begin{eqnarray}
&&A_4^{M}(r\rightarrow +\infty)= {\bar \rho}\frac{\tau_3}{2}= (\rho_M -\omega)\frac{\tau_3}{2}\nonumber\\
&&A_4^{L}(r\rightarrow +\infty)= {\bar \rho}\frac{\tau_3}{2}= (\rho_L +\omega)\frac{\tau_3}{2}
\end{eqnarray}
Thus the parameters $\rho_{L, M}$ which determine L and M dyon's actions are 
\begin{eqnarray}
&&\rho_M={\bar \rho}+\omega\nonumber\\
&&\rho_L={\bar \rho}-\omega
\end{eqnarray}
respectively. And the caloron's semi-classical free energy with dyon interactions neglected is expressed as the summation of two dyons' contribution, which reads as
\begin{eqnarray}
F^{sc}_{caloron}(T, \omega)&&=-c[|\bar\rho+\omega|^3(\frac{\Lambda}{\pi T})^{\frac{22|\bar\rho+\omega|}{6\pi T}}\nonumber\\
&&+|2\pi T-(\bar\rho-\omega)|^3(\frac{\Lambda}{\pi T})^{\frac{22|2\pi T-(\bar\rho-\omega)|}{6\pi T}}].
\end{eqnarray}
Obviously at low temperature it is minimized at $\bar\rho=\pi T$ which leads to confined Polyakov loop $Tr(L)=0$. 

A color neutral system can also be confined by anti-caloron whose two constituent anti-dyons' asymptotic behaviors are
\begin{eqnarray}
&&A_4^{\bar M}(r\rightarrow +\infty)= {\bar \rho}\frac{\tau_3}{2}= (\rho_{\bar M} +\omega)\frac{\tau_3}{2}\nonumber\\
&&A_4^{\bar L}(r\rightarrow +\infty)= {\bar \rho}\frac{\tau_3}{2}= (\rho_{\bar L} -\omega)\frac{\tau_3}{2}
\end{eqnarray}
which means anti-L and anti-M action parameters are
\begin{eqnarray}
&&\rho_{\bar M}={\bar \rho}-\omega\nonumber\\
&&\rho_{\bar L}={\bar \rho}+\omega
\end{eqnarray}
Thus the corresponding semi-classical free energy of anti-caloron reads
\begin{eqnarray}
F^{sc}_{\overline {caloron}}(T, \omega)&&=-c[|\bar\rho-\omega|^3(\frac{\Lambda}{\pi T})^{\frac{22|\bar\rho-\omega|}{6\pi T}}\nonumber\\
&&+|2\pi T-(\bar\rho+\omega)|^3(\frac{\Lambda}{\pi T})^{\frac{22|2\pi T-(\bar\rho+\omega)|}{6\pi T}}].
\end{eqnarray}
Although it is minimized at $\bar\rho=\pi T$ as well, the sign of angular velocity is different from that of caloron. This difference is induced by the CP violation in a globally rotating system. Because the caloron and anti-caloron are both neutral of color charge, a system may be composed of any ratio of the caloron and anti-caloron. A full computation by taking average of ensemble of different caloron/anti-caloron ratios will be done in next work.

For a dilute system with the same number of caloron and anti calorons, the free energy can be estimated by summing the perturbative part(Gaussian fluctuations) and semi-classical part, which is simply given by considering the action of the dyons without dyon-dyon interactions. 
\begin{eqnarray}
F_{np}(T, \omega)&&=\frac{1}{2}(F^{sc}_{ {caloron}}+F^{sc}_{\overline {caloron}}).
\end{eqnarray}
And the Polyakov loop $Tr(L)=2cos(\beta\bar\rho/2)$ or equivalently the $\bar\rho$ serves as the order parameter of the color deconfinement transition. In static case the perturbative part depends on the asymptotic value of the caloron $A_4=\rho \tau_3/2$ which reads\cite{Gross:1980br}
\begin{eqnarray}
\label{pert0}
F_{p}(T, \omega)&&=\frac{1}{3(2\pi)^2 T}{\bar \rho}^2(2\pi T-{\bar \rho})^2.
\end{eqnarray}
In a rotating system the eigen states of the quantum fluctuations around the semi-classical gauge fields involve with cylindrical Bessel functions. The momentum integration is replaced with quantum numbers corresponding to the polar radius, longitudinal coordinator and azimuthal angle which are transverse momentum $k_T=\xi_n^{(n)} R^{-1}$, longitudinal momentum $k_z$ and angular momentum number $m$, where $R$ the size of system and $\xi_n^{(m)}$ the m-th zero of the Bessel function $J_n(r)$.  The transverse momentum is discretized by considering the finite boundary condition $J_n(k_T R)=0$. And the perturbative free energy is\cite{Fujimoto:2021xix} 
\begin{eqnarray}
\label{pert}
F^\omega_{p}(T, \omega)=&&-\sum\limits_{\substack{s, m=1\\n=-\infty}}^{+\infty}\frac{e^{\frac{s n \omega}{T} }cosh(\frac{s \omega}{T})}{\pi^2 s R^3}\frac{4 \xi_n^{(m)}cos(s\frac{\bar\rho}{T})}{J_{n+1}(\xi_n^{(m)})^2}\nonumber\\
&&\times J_{n}(\xi_n^{(m)}\frac{r}{R})^2 K_1(s \frac{\xi_n^{(m)}}{T R}).
\end{eqnarray}
where $\omega$ is the real angular velocity. The $k_z$ have already been integrated out and gives the 1st order Bessel function of second kind $K_1(r)$. The remaining summations involving zeros of Bessel function do not have a nice analytical result to our knowledge. 

In the whole semi-classical calculation the angular velocity is not necessarily to be imaginary. One can easily replace the $\omega$ with $i\omega$, which corresponds to a real angular velocity in the imaginary time formalism. Although the gauge field becomes in $sl(2)$ rather than $su(2)$, it will not produce unphysical results as long as the the angular velocity is kept less than $R^{-1}$ carefully. The semi-classical free energy in the real angular velocity case is as 
\begin{eqnarray}
\label{npert}
F_{np}(T, \omega)&&=-\frac{c}{2}[sgn(\bar\rho)(\bar\rho+i \omega)^3(\frac{\Lambda}{\pi T})^{\frac{22sgn(\bar\rho)(\bar\rho+i\omega)}{6\pi T}}\nonumber\\
&&+sgn(\bar\rho_c)(\bar\rho_c+i\omega)^3(\frac{\Lambda}{\pi T})^{\frac{22sgn(\bar\rho_c)(\bar\rho_c+i\omega)}{6\pi T}}]\nonumber\\
&&-\frac{c}{2}[sgn(\bar\rho)(\bar\rho-i\omega)^3(\frac{\Lambda}{\pi T})^{\frac{22sgn(\bar\rho)(\bar\rho-i\omega)}{6\pi T}}\nonumber\\
&&+sgn(\bar\rho_c)(\bar\rho_c-i\omega)^3(\frac{\Lambda}{\pi T})^{\frac{22sgn(\bar\rho_c)(\bar\rho_c-i\omega)}{6\pi T}}]
\end{eqnarray}
where $\bar\rho_c=2\pi T-\bar\rho$ and $\omega$ now is the real angular velocity. Obviously this free energy is still real although the coefficient of angular velocity is imaginary.

\section{Critical temperature of deconfinement}
\begin{figure}[ht]
    \centering
    \includegraphics[width=8 cm]{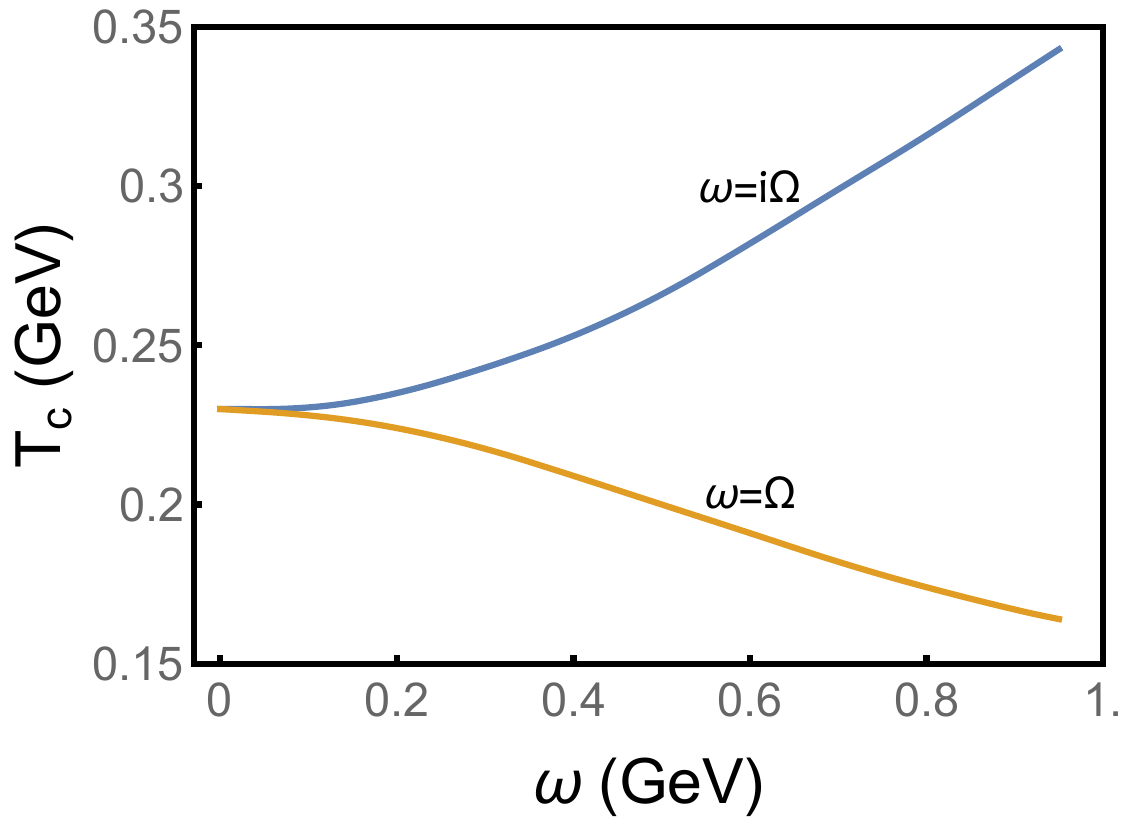}
    \caption{The critical deconfinement temperature as function of real(decrease) and imaginary(increase) angular velocity by considering the constant coupling and taking the rotational perturbative free energy.}
    \label{fig1}
\end{figure}

In order to exhibit the rotation effect more clearly we choose a relatively small radius $R=2(\pi T)^{-1}$ and $\omega=0.9 R^{-1}$. 
The critcial temperature as a function of the real angular velocity is shown in Fig. \ref{fig1}. Clearly the rotation will speed up the deconfinement transition. And the critical temperature decreases with angular speed. The imaginary velocity case is more interesting in the lattice simulation because of the sign problem. Our computation shows in Fig. \ref{fig1} that the critical temperature increases with the imaginary velocity. This is a contrary tendency comparing with the lattice result in \cite{Braguta:2021ucr, Braguta:2021jgn, Yang:2023vsw}. It is worth emphasizing that in a rotating system, the perturbative free energy is modified by not only the rotation polarization effect but also the finite size condition. This means this part of the free energy is different from that in static case even $\omega=0$. Nonetheless, the perturbative free energy is still unable to confine the system.

\begin{figure}[ht]
    \centering
    \includegraphics[width=8 cm]{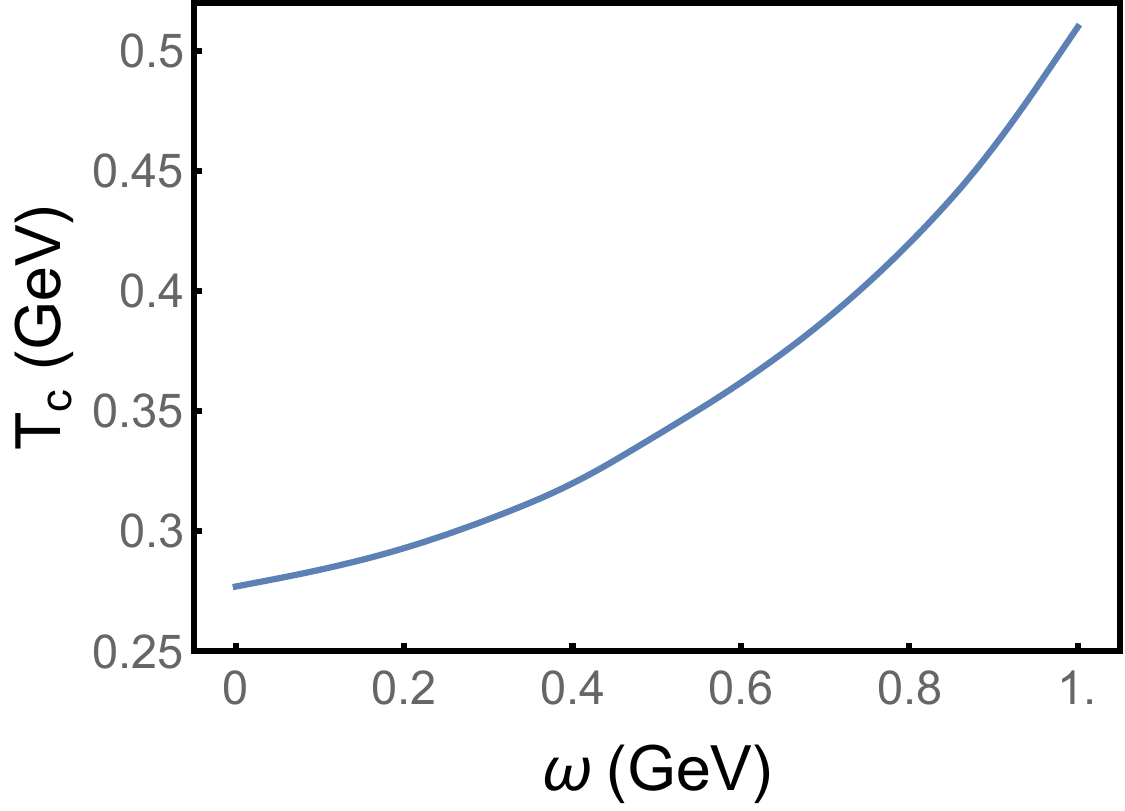}
    \caption{The critical deconfinement temperature as function of angular speed by considering the coupling running and taking the static perturbative free energy.}
    \label{fig2}
\end{figure}

\begin{figure}[ht]
    \centering
    \includegraphics[width=8 cm]{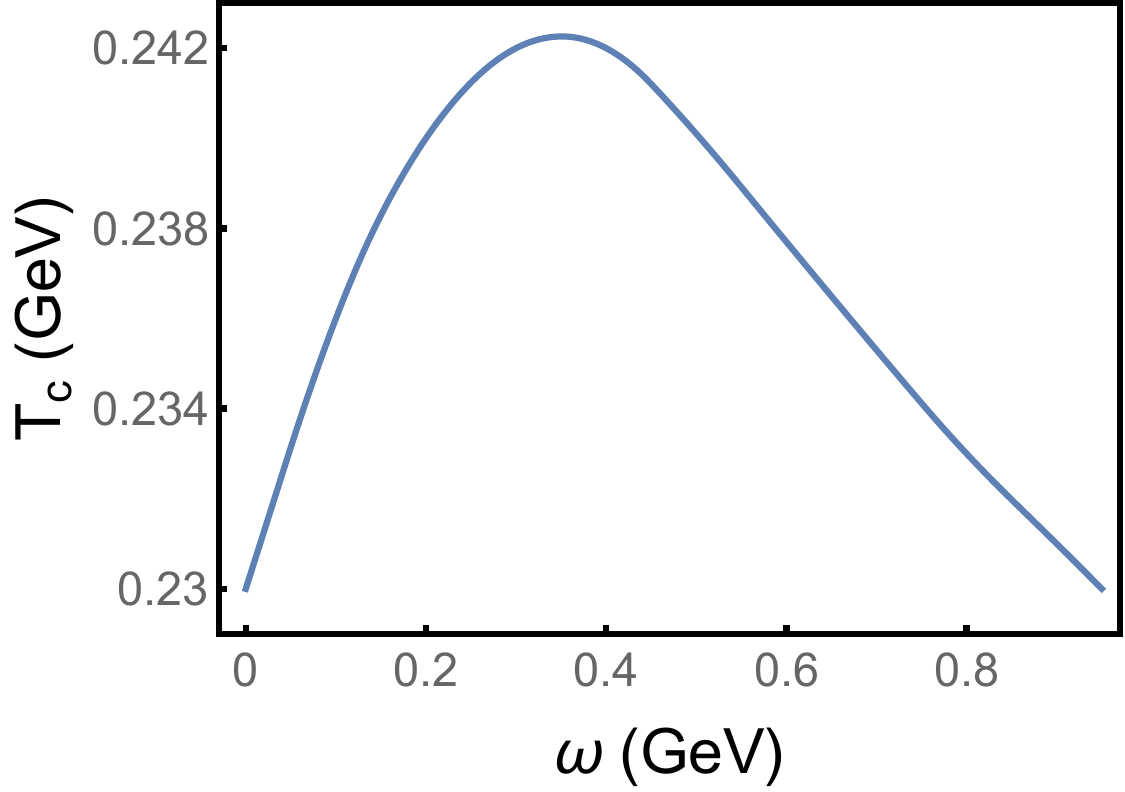}
    \caption{The critical deconfinement temperature as function of angular speed by considering the coupling running and taking the rotational perturbative free energy.}
    \label{fig3}
\end{figure}

In order to understand the lattice result, one should notice that there is another important fact which has been ignored in the computation. That is the coupling constant is suggested to increase with the real angular velocity in \cite{Jiang:2021izj}. In order to show this effect we introduce a rotation-dependent coupling constant $g(\omega)=(1+0.1 |\omega|/\Lambda)g$ 
to magnify the non-perturbative part according to Eq.\ref{lam} and use the perturbative free energy in static case to show the pure coupling running consequence. Combining the results in Eq.\ref{pert0} and Eq.\ref{npert}, the critical temperature increases with the real angular velocity in Fig. \ref{fig2} now. It is understandable because a larger coupling will surely bound color charges more tightly and thus make the confinement harder to be broken. This effect can be directly checked by noticing the confined potential, which is given by the semi-classical part, is deeper when the coupling is stronger. Because there is no good analytical result for the $\omega$ dependence of coupling constant, it is not a straightforward computation to continue the angular velocity to a imaginary one. Different $\omega$ form, such as $|\omega|$ or $\omega^2$ will give very different results in the imaginary velicity case. This is the reason why no imaginary result shown in Fig.\ref{fig2} and \ref{fig3}.

In the above computation it is clear that there is only one factor help to confine color charges, that is a stronger coupling constant. While both the semi-classical gluon vacuum and the corresponding Gaussian fluctuation penalize the confinement. Hence one can expect that a competition between these two part may lead to a non monotonic tendency of critical temperature as the function of angular velocity. 
Taking both the running coupling and rotational perturbative free energy into account, the total free energy density, i.e. Eq.\ref{pert} and Eq.\ref{npert}, gives the critical temperature in Fig. \ref{fig3}.

As expected a non-monotonic behavior is obtained and can be understood by the competition between two effects in above. When rotation is slow, the running coupling dominates because the coupling dependence is in the action of dyon according to Eq.\ref{lam}. This slows down the deconfinement process. As rotation increases, the angular velocity serves as a chemical potential in the perturbative part and changes the weight of dyon. This helps the color deconfinement and make the critical temperature decreases with rotation. In Fig. \ref{fig3} it shows the increasing part is not quite significant with the chosen set of parameters. 

\section{Conclusions}
In this work we have reformulated the non-abelian gauge theory in a local inertial frame with the presence of a background rotation. Starting from the rotation-modified Yang-Mills Lagrangian density, the new gauge transformation and Polyakov loop have been established. With this new formalism we have studied the influence of background rotation on the color deconfinement transition for a SU(2) gauge symmetric system. The KvBLL caloron, which is a color neutral and asymptotically nontrivial solution of Yang-Mills equation at finite temperature, is adopted to confine the color charges. New solutions of dyons, which serve as constituent particles of caloron, have been obtained. These new dyon configurations only change the $A_4$ component by a shifting $\omega$ but leave the spatial parts and corresponding actions invariant. This indicates that the constituted caloron may be compounded in the same after taking the $A_4$ shifting of L and M dyons into account as Fig. \ref{fig0}. And in the region far away from the core of dyons, this constituent pattern has also been confirmed analytically. As a result, the semi-classical potential of such rotating caloron system, which confines color charges, have been obtained in the same way as the static case. Combining the semi-classical potential with the perturbative part induced by the Gaussian fluctuation around the background caloron, the color deconefinement transition have been studied by computing the free energy of the system at different temperatures. To determine which one of the three main facts, i.e. the impacts of semi-classical part, perturbative part and running coupling of rotation, will lead to a non-trivial tendency of the critical temperature as a function angular velocity, different computation schemes have been discussed. It is found that neither the rotational semi-classical potential nor Gaussian fluctuations can confine color charges more tightly when the rotation becomes faster. Only a stronger coupling constant is able to make the critical temperature increasing with angular velocity, as that indicated in lattice simulations. And it is also found with some particular sets of parameters, a non-monotonic dependence of the critical temperature will be obtained in the most physically realistic case, in which all the three factors are taken into account. 

This work is expected to be improved from three aspects. First is the caloron part. One could go beyond dilute system limit and consider the center of dyon locating at arbitrary positions to obtain an explicitly analytical KvBLL caloron under rotation. In the arbitrary center location case the Eq.\ref{a4} becomes more complicated and may introduce extra terms into spatial components of gauge field. Second, the running coupling constant clearly plays an key role in the non-trivial dependence of critical temperature on angular velocity. A independent computation from the work in \cite{Jiang:2021izj} with more standard and systematic method is necessary to confirm the running behavior of the strong coupling constant with the angular velocity. Third the quarks/fermions are not taken into account in this work. The lattice simulation \cite{Braguta:2021ucr} indicates the fermions may provide a contrary contribution to the gluons'. This can be checked in our framework for the real velocity case as well.

{\bf Acknowledgments.}
The work of this research is supported by the National Natural Science Foundation of China, Grant Nos. 12375131(YJ). We thanks Xu-Guang Huang very much for his invaluable comments.

\end{document}